\documentclass[aps,preprint,amsmath,amssymb,amsfonts,nofootinbib]{revtex4-2}
\pdfoutput=1

\usepackage{epsfig}
\usepackage{graphicx}
\usepackage{dcolumn}
\usepackage{bm}
\usepackage{amsthm}
\usepackage{amsmath}
\usepackage{color}
\usepackage[allcolors=blue,colorlinks=true]{hyperref}
\usepackage{xspace}
\usepackage{tikz}
\usetikzlibrary{decorations.pathmorphing, arrows.meta, calc, patterns, patterns.meta}
\usepackage[caption=false]{subfig}

\newcommand{\p}{\partial}

\newcommand{\dd}{\mathrm{d}}

\newcommand{\pdU}{\partial_U}
\newcommand{\pdV}{\partial_V}

\begin{document}
	
\preprint{CCTP-2025-18, ITCP-IPP 2025/18}

\title{Approaching a dynamical extreme black hole horizon}
%
\author{Achilleas P. Porfyriadis}
\email{porfyriadis@physics.uoc.gr}
\author{Christopher Rosen}
\email{rosen@physics.uoc.gr}
\author{Georgios Tsaraktsidis}
\email{tsaraktsidisg@gmail.com}
\affiliation{Crete Center for Theoretical Physics, Institute of Theoretical and Computational Physics,\\
	Department of Physics, University of Crete, 70013 Heraklion, Greece}

\date{\today}
\begin{abstract}

We give an explicit closed form description of the late-time near-horizon approach to dynamical extreme Reissner–Nordström (DERN) black holes. 
These are spherically symmetric dynamical solutions of Einstein-Maxwell theory coupled to a neutral scalar that feature: (i) a spacetime metric which tends to that of a static extreme Reissner–Nordström (RN), and (ii) a scalar field which exhibits the linear Aretakis instability ad infinitum in the non-linear theory.
We employ the two-dimensional Jackiw-Teitelboim (JT) gravity to solve explicitly for the non-linear s-wave dynamics of the four-dimensional theory near an ${\rm AdS}_2\times {\rm S}^2$ throat. 
For a teleologically defined black hole horizon, we impose boundary conditions on JT's dilaton field (which encodes the gravitational dynamics) and the scalar matter as follows: (i) the JT dilaton decays at late times on the ${\rm AdS}_2$ boundary to a value that corresponds to a static extreme RN in the exterior, and (ii) the scalar obeys boundary conditions characteristic of linear Aretakis behavior on ${\rm AdS}_2$.
We ensure our DERN solutions are singularity-free and we note that our approach to DERN is accompanied by a final burst of outgoing scalar matter flux leaking out of the ${\rm AdS}_2$ throat. 
The boundary conditions we impose on the JT dilaton place its late-time boundary profile on the threshold of black hole formation with sub-extreme and super-extreme RN on either side of our DERNs.

\end{abstract}


\maketitle\newpage

\tableofcontents

\section{Introduction}

Extreme and near-extreme black holes have recently become a focal point of research in theoretical high-energy physics on the one hand and mathematical and numerical relativity on the other. 
This comes at a time when astronomers keep measuring nearly extreme spin values in observations of rotating black holes using electromagnetic \cite{Reynolds:2020jwt} and gravitational \cite{LIGOScientific:2025rsn} waves.

In the high-energy physics literature, extreme black holes are currently subject to intense investigation due to the realization that there are large quantum mechanical fluctuations in their ${\rm AdS}_2$ throat region, which are governed by solvable effective two-dimensional models of gravity (see e.g. the review \cite{Mertens:2022irh}). This has significant implications for a variety of aspects concerning near-extreme black holes, including their thermodynamics \cite{Iliesiu:2020qvm, Iliesiu:2022onk} and Hawking evaporation \cite{Brown:2024ajk}.
The core model of two-dimensional gravity in these developments is the Jackiw-Teitelboim (JT) theory \cite{Jackiw:1984je, Teitelboim:1983ux}. Within spherical symmetry, to a large degree of universality, JT describes the flow of any higher dimensional gravity to ${\rm AdS}_2$, and provides an exactly solvable theory of the non-linear gravitational dynamics in the ${\rm AdS}_2$ throats of extreme black holes \cite{Almheiri:2014cka,Almheiri:2016fws,Maldacena:2016upp,Nayak:2018qej}.\footnote{That said, there are aspects of backreaction in ${\rm AdS}_2$ which are not captured by JT \cite{Castro:2025pst}. Moreover, efforts to utilize JT for rotating extreme black holes are fraught with substantial complexities \cite{Castro:2021csm}.}

In the mathematical and numerical relativity literature, extreme black holes are also currently a source of profound fascination due to the discovery that they are characterized by weak linear instabilities which allow for conjectures and proofs of non-linear stability with features of weak instability (see e.g. the perspective \cite{Dafermos:2025int}).
Additionally, in opposition to the third law of black hole mechanics, it has been rigorously established that an extreme Reissner–Nordström event horizon may be formed in finite time from regular gravitational collapse of a charged scalar, in such a way that at earlier times the black hole has a non-extreme apparent horizon \cite{Kehle:2022uvc}.

The harbinger of weak instabilities associated with extreme black holes was the celebrated Aretakis instability \cite{Aretakis1, Aretakis2} for linear perturbations. Following that, the pioneering numerical study in \cite{Murata:2013daa} argued that there exist dynamical extreme Reissner–Nordström (DERN) black holes. They are the focus of the present work.
DERNs are spherically symmetric solutions in the non-linear theory of Einstein-Maxwell coupled to a neutral scalar, whose metric tends to that of the static extreme Reissner–Nordström (ERN), while the scalar exhibits the Aretakis instability ad infinitum.\footnote{Recently it was shown that for a planar extreme horizon, with infinite area, such asymptotic behavior is the generic non-linear fate of the linear Aretakis instability \cite{Horowitz:2025ayc}.}
It was recently established rigorously that, within spherical symmetry, DERN is stable in the non-linear theory \cite{Angelopoulos:2024yev}. Moreover, it is argued numerically in \cite{Murata:2013daa} and conjectured mathematically in \cite{Angelopoulos:2024yev} that DERN lies on a threshold of black hole formation from critical collapse in this theory. This is a different threshold from the familiar Choptuik threshold \cite{Choptuik:1992jv} of naked singularities. 

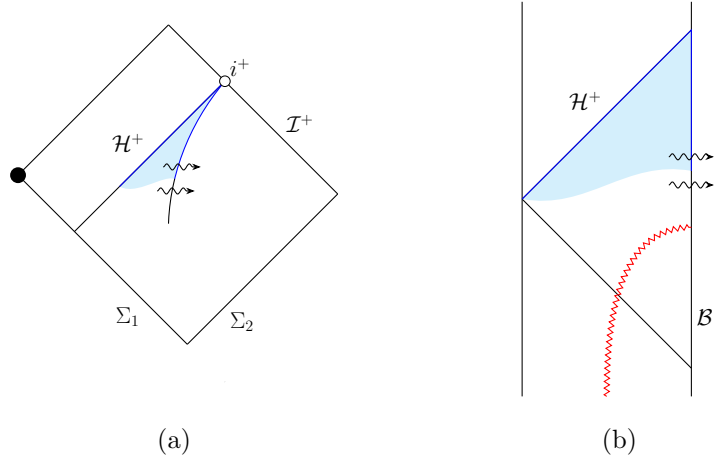
\begin{figure}[h]
	\centering
	\subfloat[]{\scalebox{0.5}{
			\begin{tikzpicture}[scale=4]
				\fill[cyan!30, opacity=0.5] (1.5,-0.5) to [out=130, in=-130] (1.5, 1.5) -- (0.5,0.5) -- (1.5, -0.5);
				\draw[blue, thick] (1.5,-0.5) to [out=130, in=-130] (1.5, 1.5);
				\fill[white] (0.8, 0.8) to [out=-30, in=100] (1.2,0.8) -- (1.6, -0.6) -- (1.4,-0.6) -- (0.5, 0.5) -- (1, 1);
				\draw [thick, dash pattern= on 0 off 130pt on 34.35pt off 200pt]
				(1.5,-0.5) to [out=130, in=-130] (1.5, 1.5);
				\draw[thick] (1.5,1.5) -- (2.25,0.75) node[midway, above right] {\Large $\mathcal{I}^+$};
				\draw[thick] (1.5,1.5) -- (0.5,0.5) node[midway, above left] {\Large $\mathcal{H}^+$};
				\draw[thick] (1.25,-0.25) -- (2.25,0.75) node[pos=0.25, below right] {\Large $\Sigma_2$};
				\draw[thick] (1.25,-0.25) -- (0.125,0.875) node[pos=0.25, below left] {\Large $\Sigma_1$};
				\draw[thick] (0.125, 0.875) -- (1.125,1.875);
				\draw[thick] (1.5, 1.5) -- (1.125,1.875); 
				\draw[blue, thick] (1.5, 1.5) -- (0.8, 0.8);
				\draw[fill=white] (1.5,1.5) circle(1pt) node[above right] {\Large $i^+$};
				\fill[black] (0.125,0.875) circle(1.5pt);
				\tikzset{radiation/.style={
						decorate, 
						decoration={snake, amplitude=0.8mm, segment length=3mm, post length=1mm},
						-Stealth, 
						thick
				}}
				\draw[radiation] (1.09,0.93) -- (1.34,0.93);
				\draw[radiation] (1.05,0.77) -- (1.3,0.77);
			\end{tikzpicture}
		}\label{Fig:mainDERN}}
	\qquad\qquad\qquad
	\subfloat[]{\scalebox{0.5}{
			\begin{tikzpicture}[scale=1.5]
				\coordinate (Vertex) at (0,0);
				\coordinate (TopRight) at (3, 3);
				\coordinate (BotRight) at (3, -3);
				\draw[thick, black] (0,3.5) -- (0,-3.5) ;
				\draw[thick, black] (3,3.5) -- (3,-3.5) node[pos=0.8, right] {\Large $\mathcal{B}$} ;
				\draw[thick] (Vertex) -- (TopRight) node[midway, above left] {\Large $\mathcal{H}^+$};
				\draw[thick] (BotRight) -- (Vertex) ;
				\fill[cyan!30, opacity=0.5] (Vertex) to [out=-10, in=170] (3,0.5) -- (TopRight) -- (Vertex);
				\draw[thick, blue] (Vertex) -- (TopRight) -- (3,0.5);
				\draw[decorate, decoration={zigzag, amplitude=2pt, segment length=5pt}, thick, red] (1.5, -3.5) to [out=90, in=180] (3, -0.5);;
				\tikzset{radiation/.style={
						decorate, 
						decoration={snake, amplitude=0.8mm, segment length=3mm, post length=1mm},
						-Stealth, 
						thick
				}}
				\draw[radiation] (2.6,0.75) -- (3.4,0.75);
				\draw[radiation] (2.6,0.25) -- (3.4,0.25);
				
			\end{tikzpicture}
		}\label{Fig:mainAdS}}
	\caption{Penrose diagrams showing: (a) DERN arising from maximal development of characteristic data on $\Sigma_1\cup\Sigma_2$ (the data on $\Sigma_1$ ends in the solid point where it is incomplete). The shaded blue region at late times near the horizon is given by a dynamical ${\rm AdS}_2\times {\rm S}^2$ throat. (b) The ${\rm AdS}_2\times {\rm S}^2$ throat with dynamics described by the JT theory. A Poincare horizon is identified with the black hole horizon $\mathcal{H}^+$ and appropriate boundary conditions are imposed on the dilaton and matter fields in JT so as to obtain a solution in the shaded blue region that matches DERN. There is matter flux leaking out of ${\rm AdS}_2\times {\rm S}^2$.
	}\label{Fig:main}
\end{figure}

In this paper, we study analytically the late-time near-horizon approach to DERN using the JT description of its gravitational dynamics (see Fig~\ref{Fig:main}). 
We begin by reviewing in Section~\ref{Section2} the linear Aretakis behavior of a neutral scalar perturbation on a fixed ERN and its non-linear fate leading to DERN in the full Einstein-Maxwell-Scalar theory.
In Section~\ref{Section3}, we go over the emergence of ${\rm AdS}_2\times {\rm S}^2$ from ERN as well as sub-extreme RN and super-extreme RN. This helps clarify the critical nature of the boundary condition we seek to impose on the dilaton field in JT. We also review the boundary conditions that a scalar in ${\rm AdS}_2$ needs to obey in order to display linear Aretakis behavior with respect to a preferred Poincare horizon. 
In Section~\ref{Sec:DERNinJT} we solve the JT equations and impose the boundary conditions corresponding to a DERN. We make sure to arrange for sufficient scalar matter flux to leak through the ${\rm AdS}_2$ boundary so as to maintain the validity of the JT theory and a singularity-free horizon.

\section{Aretakis instability and DERN}\label{Section2}

In this section we review what is known about the linear Aretakis instability and its fate at the non-linear level. 
We focus on the dynamical extreme Reissner–Nordström (DERN) solutions which are the subject of this paper.

Our conventions in this paper are such that Reissner–Nordström (RN) takes the form:\footnote{Our units are geometrized with $G=c=1$ and while most of the literature considers an electrically charged RN, we find it slightly more convenient to work with the magnetic solution.}
\begin{align}\label{RN}
	\begin{aligned}
		{\rm d}s^2&=-\left(1-\frac{2M}{\hat{r}}+\frac{Q^2}{\hat{r}^2}\right){\rm d}\hat{t}^2+
		\left(1-\frac{2M}{\hat{r}}+\frac{Q^2}{\hat{r}^2}\right)^{-1}{\rm d}\hat{r}^2+
		\hat{r}^2 {\rm d}\Omega^2 \\
		F&=Q\sin\theta \, {\rm d}\theta\wedge {\rm d}\phi.
	\end{aligned}
\end{align}
Extreme RN (ERN) is defined by $Q=M$ and its event horizon lies at $\hat{r}=M$.

\subsection{Linear Aretakis instability on ERN}\label{Sec:LinearAretakisERN}

When Aretakis set out to explore the stability of ERN under linear perturbations by a massless neutral scalar field $\phi$, he discovered that while $\phi$ itself does indeed decay at late times everywhere on and outside the future horizon $\mathcal{H}^+$, its transverse derivatives, $\p_\perp^n \phi$, do not all decay at late times on the horizon \cite{Aretakis1, Aretakis2}. Specifically, for generic initial data, the $\ell$'th harmonic $\phi_\ell$ has its first $\ell$ transverse derivatives decay, $\p_\perp^k \phi\vert_{\mathcal{H}^+}\to 0$ for $k\leq l$, but the next derivative approaches a constant $\p_\perp^{l+1} \phi\vert_{\mathcal{H}^+}\to H_\ell$ and higher derivatives blow up beginning with $\p_\perp^{\ell+2} \phi\vert_{\mathcal{H}^+}\sim H_\ell \, v$, where $v$ is the Killing time along $\mathcal{H}^+$. Here $H_\ell$, called an Aretakis constant, is non-vanishing for generic initial data with support on $\mathcal{H}^+$. For scattering data, supported away from the horizon, the instability is weaker in the sense that one needs an extra derivative to get to blow up, but otherwise still present \cite{LMRT1212, Aretakis4}. Precise late-time asymptotics on and outside $\mathcal{H}^+$, including the leading-order coefficients in terms of initial data, have been given in \cite{LMRT1212, Angelopoulos:2018uwb}.

A plethora of analytic and numerical work in the physics and mathematics literature has established the linear Aretakis instability as a robust dynamical feature associated with any perturbation of an extreme black hole horizon.  
This includes massive and charged scalars as well as (coupled) gravitational and electromagnetic perturbations, of extreme RN as well as Kerr black holes, and in four as well as higher dimensions \cite{LMRT1212, Apetroaie:2022rew, Lucietti:2012sf, Murata:2012ct, Aretakis:2012ei, Casals:2016mel, Gajic:2023uwh, Zimmerman:2016qtn, Gelles:2025gxi}.

In this paper we focus on the case of a spherically symmetric massless neutral scalar. 
As a linear perturbation on a fixed ERN in Eq.~\eqref{RN}, the scalar, $\phi$, may be taken to satisfy the wave equation $\Box\phi=0$. 
Using the standard ingoing Eddington-Finkelstein coordinate $\hat{v}=\hat{t}+\hat{r}_*$, solutions $\phi=\phi(\hat{v},\hat{r})$ to the scalar wave equation have been shown to exhibit the following behavior on and outside the extreme horizon $\hat{r}=M$:
\\
\textbullet\ \textit{Non-vanishing Aretakis constant}: $ H\neq 0$
\begin{align}\label{linear Aretakis H}
	\begin{aligned}
		&\phi\vert_{\hat{r}=M}\sim -\frac{2H}{\hat{v}}\,, \quad \phi\vert_{\hat{r}=\hat{r}_0>M}\sim -\frac{4H}{(\hat{r}_0-M) \hat{v}^2}\\
		&\p_{\hat{r}}\phi\vert_{\hat{r}=M}\to H\,, \quad \p_{\hat{r}}^2\phi\vert_{\hat{r}=M}\sim -H \hat{v}
	\end{aligned}
\end{align}
\textbullet\ \textit{Vanishing Aretakis constant}: $H=0$
\begin{align}\label{linear Aretakis no H}
\begin{aligned}
	&\phi\vert_{\hat{r}=M}\sim \frac{1}{\hat{v}^2}\,, \quad \phi\vert_{\hat{r}=\hat{r}_0>M}\sim \frac{1}{\hat{v}^3}\\
	&\p_{\hat{r}}\phi\vert_{\hat{r}=M}\to 0\,, \quad \p_{\hat{r}}^2\phi\vert_{\hat{r}=M}\to \text{cnst}\,, \quad \p_{\hat{r}}^3\phi\vert_{\hat{r}=M}\sim \hat{v}
\end{aligned}
\end{align}
The above instabilities of the second or third transverse derivatives of solutions to the linear wave equation, naturally raised the question of what is their fate in the full non-linear theory of Einstein-Maxwell coupled to a neutral scalar.

\subsection{The nonlinear fate of the instability and DERN}\label{Sec:DERN}

In the non-linear Einstein-Maxwell-Scalar theory,
\begin{align}\label{EMS}
	\begin{gathered}
		R_{\mu\nu}-\frac{1}{2}R\, g_{\mu\nu} = 2\left(F_{\mu\rho}F_\nu\,^\rho -\frac{1}{4} g_{\mu\nu} F^2\right) +\nabla_\mu\phi\nabla_\nu\phi -\frac{1}{2}g_{\mu\nu}(\nabla\phi)^2 \\
		{\rm d}F={\rm d}\star F = 0 \,, \qquad \Box\phi = 0
	\end{gathered}
\end{align}
gravitational backreaction has been shown numerically to regulate the instability by turning it into a transient one. Specifically, Ref.~\cite{Murata:2013daa} showed that, generically, backreacting a spherically symmetric massless neutral scalar on extreme RN produces a near-extreme RN with the transverse derivatives of the scalar growing on its horizon for a timescale inversely proportional to its surface gravity, before eventually decaying.
More precisely, \cite{Murata:2013daa} considered spherically symmetric characteristic initial data on a pair of bifurcate null hypersurfaces $\Sigma_1\cup\Sigma_2$ such that the spacetime $(M_i,Q)$ and scalar $\phi\sim\mathcal{O}(\epsilon)$ data give rise to a future event horizon $\mathcal{H}^+$ which is the ERN one when $\epsilon=0$ and $M_i=Q$. Here, $Q$ is the Maxwell charge, which is conserved in the theory \eqref{EMS}, and $M_i$ is the initial Bondi mass. 
Two kinds of data, whose linearization yields a perturbative test scalar evolving on a fixed ERN (as described in the previous subsection), were considered: $M_i=Q+\mathcal{O}(\epsilon^2)$ and $M_i=Q+\mathcal{O}(\epsilon)$.\footnote{The scalar's stress tensor couples to the metric at quadratic order, so a first order metric perturbation with $M_i=Q+\mathcal{O}(\epsilon)$ simply increases the RN mass but doesn't couple to the scalar at leading order.}
For generic initial data of the first kind, $M_i=Q+\mathcal{O}(\epsilon^2)$, compactly supported near $\mathcal{H}^+$, the numerical evolution settles down to a near-extreme RN  with surface gravity $\kappa=\mathcal{O}(\epsilon)$. 
Prior to that, for a long time of order $\mathcal{O}(1/\epsilon)$, the transverse derivatives of the scalar behave as in the linear Aretakis story of the previous subsection: $\p_\perp \phi\vert_{\mathcal{H}^+}$ remains approximately constant for $\mathcal{O}(1/\epsilon)$ time (before beginning a slow decay $\sim e^{-\kappa v}$), while $\p_\perp^2 \phi\vert_{\mathcal{H}^+}$ grows for $\mathcal{O}(1/\epsilon)$ time to a \emph{finite} maximum value (before beginning its slow decay): $\lim_{\epsilon\to 0}\,  \textrm{max}\left(\p_\perp^2 \phi\vert_{\mathcal{H}^+}\right)\neq 0$. This means that the Aretakis instability survives in the non-linear theory, albeit, generically only as a \emph{transient} phenomenon.
For data of the second kind, $M_i=Q+\mathcal{O}(\epsilon)$, or for scattering data, supported away from $\mathcal{H}^+$, similar but weaker behavior was observed, namely the final RN black hole has $\kappa=\mathcal{O}(\epsilon^{1/2})$ and one needs an extra derivative to get to the finite growth on the horizon: $\lim_{\epsilon\to 0}\,  \textrm{max}\left(\p_\perp^3 \phi\vert_{\mathcal{H}^+}\right)\neq 0$.
An account of generic transient Aretakis instability using the JT theory was given in \cite{Hadar:2018izi}.
 
On the other hand, Ref.~\cite{Murata:2013daa} also argued that there exist 
dynamical extreme RN (DERN) black holes: fine-tuned non-linear dynamical solutions which remain extremal and exhibit the Aretakis instability indefinitely. 
DERNs have a regular extreme event horizon $\mathcal{H}^+$, with extremality defined e.g. by the absence of trapped surfaces, and
are characterized by the following two facts: (i) the spacetime metric asymptotes, on and outside $\mathcal{H}^+$, to the static ERN one, and (ii) the scalar's asymptotic behavior, on and outside $\mathcal{H}^+$, matches the linear behavior of a test field as summarized in \eqref{linear Aretakis H}. This of course means that, in the coupled theory \eqref{EMS}, certain components of the Ricci tensor for DERN behave analogously to \eqref{linear Aretakis H}. Taken together, these facts imply that DERN is a solution which at late times has a  dynamical extreme horizon glued onto a static ERN exterior. 
In this paper, we will realize this configuration as a solution of the JT equations with suitable boundary conditions.

Numerically, DERN was found to lie at a critical value $M_*(\epsilon)$ which delineates the space of initial data considered in \cite{Murata:2013daa}, such that $M_i>M_*(\epsilon)$ gives rise to a sub-extreme RN black hole, while $M_i<M_*(\epsilon)$ produces no event horizon, thereby giving rise to  super-extreme RN. 
Our boundary conditions for obtaining DERN in JT theory will ensure this criticality.
It is worth emphasizing that the super-extreme RN spacetime here does not have a naked singularity as this is dynamically inaccessible: in Fig.~\ref{Fig:mainDERN} the data on $\Sigma_1$ ends in the solid point where it is incomplete. 
The theory \eqref{EMS} contains only a neutral scalar and so the Maxwell charge $Q$ is conserved. As a result, DERN lies at the threshold of black hole formation but because charge cannot be radiated away in this theory, super-extreme RN replaces a solution that is expected to disperse in a theory containing charged matter.\footnote{For example, extreme black holes have been rigorously shown to occur at the threshold between dispersion to Minkowski and a sub-extreme BH in Einstein-Maxwell-Vlasov theory \cite{Kehle:2024vyt}.
A first step towards extension of the analysis of \cite{Murata:2013daa} to the case of charged scalar critical collapse was recently taken in \cite{Gelles:2025gxi}.}

While the above means that DERN is fine-tuned, it was recently shown rigorously \cite{Angelopoulos:2024yev} that, within spherical symmetry, it is also stable in the following sense: there exists a codimension-1 submanifold $\mathfrak{M}_\text{stable}$ of the moduli space of initial data $\mathfrak{M}$ that are close to ERN, such that all data in $\mathfrak{M}_\text{stable}$ evolve to DERN. It is expected then that $\mathfrak{M}_\text{stable}$ delineates the boundary between black hole formation and dispersion (within the domain of dependence of $\Sigma_1\cup\Sigma_2$).

\section{Emergence of ${\rm AdS}_2$ and linear Aretakis on it}\label{Section3}

In this section we go over the emergence of ${\rm AdS}_2$ from three limits of RN spacetimes near extremality.
We also  review the ${\rm AdS}_2$ analysis appropriate for identifying linear Aretakis instability of scalar field perturbations on ERN black holes.

\subsection{${\rm AdS}_2$ from three limits of RN}\label{Sec:ThreeLimits}

There are three physically distinct scalings of RN spacetimes which all yield an ${\rm AdS}_2\times {\rm S}^2$ in the limit, each coming with a distinct first order correction. They correspond to taking the limit from extreme, sub-extreme, and  super-extreme RN (ERN, sub-ERN, and super-ERN, respectively). 
Below we go over the three limits following closely Ref.~\cite{Hadar:2020kry}, which explained in detail the first two cases. The third case, presented here, helps identify the criticality of the boundary conditions imposed in Section \ref{Sec:DERNinJT} for the desired DERN solution.

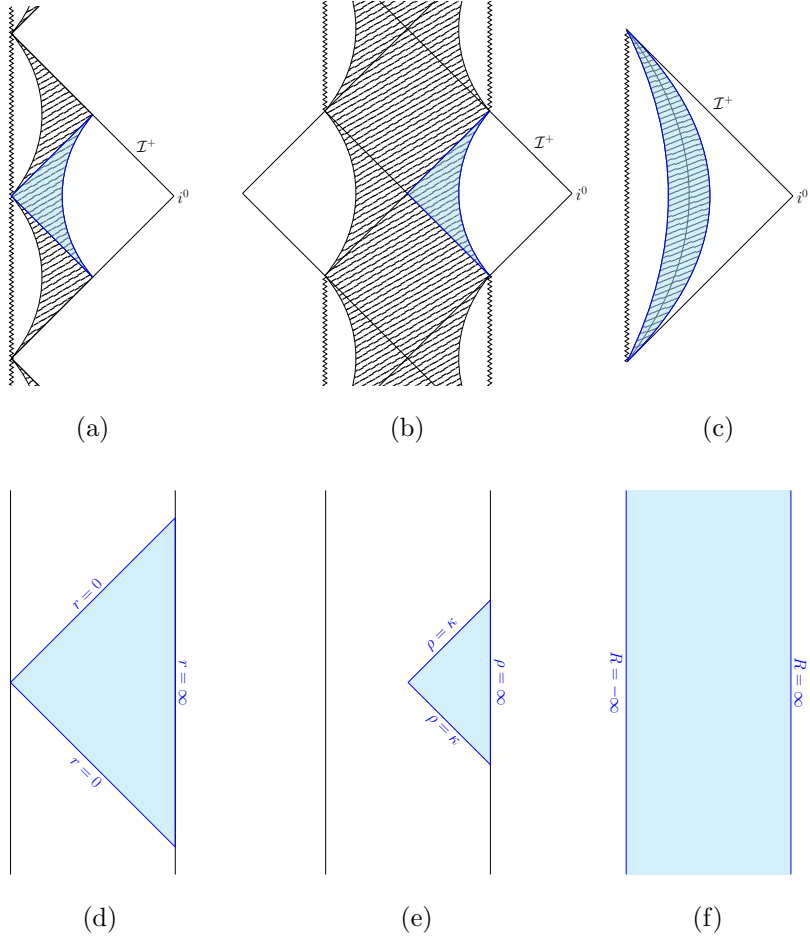
\begin{figure}[h]
	\centering
	\subfloat[]{\scalebox{0.36}{
			\begin{tikzpicture}[scale=2]
				\clip (-0.3, -3.5) rectangle (3.3, 3.5);
				\coordinate (top) at (0, 3);
				\coordinate (bottom) at (0, -3);
				\coordinate (right) at (3, 0);
				\coordinate (center) at (0, 0);
				\draw[thick, decorate, decoration={zigzag, amplitude=2pt, segment length=5pt}] (0,-3.5) -- (0,3.5);
				\draw[thick] (top) -- (right) 
				node[pos=0.75, above right] {\Large$\mathcal{I}^+$};
				\draw[thick] (bottom) -- (right);
				\draw[thick] (center) -- (3, 3);
				\draw[thick] (3, 3) -- (1.5, 4.5);
				\draw[thick] (center) -- (3, -3);
				\draw[thick] (3, -3) -- (1.5, -4.5);
				\draw[thick] (0, 3) -- (1.5, 4.5);
				\draw[thick] (0, -3) -- (1.5, -4.5);
				\draw[pattern={Lines[angle=30, distance=2pt]}]
				(1.5,-4.5) to [out=130, in=-130] (1.5, -1.5) to [out=130, in=-130]
				(1.5, 1.5) to [out=130, in=-130] (1.5, 4.5)
				--
				(top) -- (1.5, 4.5) -- (0, 6) to [out=-50, in=50] (0, 3)
				to [out=-50, in=50] (center) to [out=-50, in=50] (bottom) to [out=-50,
				in=50] (0, -6) -- (1.5, -4.5) -- (bottom)
				-- cycle;
				\fill[cyan!30, opacity=0.5] (1.5,-1.5) to [out=130, in=-130] (1.5, 1.5)
				-- (center) -- (1.5, -1.5);
				\draw[blue,thick] (1.5,-1.5) to [out=130, in=-130] (1.5, 1.5) --
				(center) -- (1.5, -1.5);
				\node [right] at (3,0) {\Large$i^0$};
			\end{tikzpicture}
	}}
	\quad
	\subfloat[]{\scalebox{0.365}{
			\begin{tikzpicture}[scale=2]
				\clip (-3.3, -3.5) rectangle (3.3, 3.5);
				\draw[thick] (-1.5, 4.5) -- (3, 0) node[pos=0.833, above right] {\Large$\mathcal{I}^+$};
				\draw[thick] (1.5, 4.5) -- (-3,0);
				\draw[thick] (-3, 0) -- (1.5, -4.5);
				\draw[thick] (3, 0) -- (-1.5, -4.5);
				\draw[thick] (-1.5, 1.5) -- (1.5, -1.5);
				\draw[thick] (1.5, 1.5) -- (-1.5, -1.5);
				\draw[decorate, decoration={zigzag, amplitude=2pt, segment length=5pt}, thick] (-1.5, 4.5) -- (-1.5, 1.5);
				\draw[decorate, decoration={zigzag, amplitude=2pt, segment length=5pt}, thick] (1.5, 4.5) -- (1.5, 1.5);
				\draw[decorate, decoration={zigzag, amplitude=2pt, segment length=5pt}, thick] (-1.5, -1.5) -- (-1.5, -4.5);
				\draw[decorate, decoration={zigzag, amplitude=2pt, segment length=5pt}, thick] (1.5, -1.5) -- (1.5, -4.5);
				\draw[pattern={Lines[angle=30, distance=2pt]}]  
				(1.5, -4.5) to [out=130, in=-130]  (1.5, -1.5) to [out=130, in=-130] (1.5, 1.5) to [out=130, in=-130] (1.5, 4.5)
				-- 
				(-1.5, 4.5) 
				to [out=-50, in=50] (-1.5, 1.5) to [out=-50, in=50] (-1.5, 1.5) to [out=-50, in=50] (-1.5, -1.5) to [out=-50, in=50] (-1.5, -4.5)
				--
				(1.5, -4.5)
				-- cycle;
				\fill[cyan!30, opacity=0.5] (1.5, -1.5) to [out=130, in=-130] (1.5, 1.5) -- (0,0) -- (1.5,-1.5);
				\draw[blue, thick] (1.5, -1.5) to [out=130, in=-130] (1.5, 1.5) -- (0,0) -- (1.5,-1.5);
				\node [right] at (3,0) {\Large$i^0$};
			\end{tikzpicture}
	}}
	\quad
	\subfloat[]{\raisebox{0.8em}{\scalebox{0.368}{
				\begin{tikzpicture}[scale=2]
					\coordinate (bottom) at (0,-3);
					\coordinate (top) at (0,3);
					\coordinate (i0) at (3,0);
					\draw[black] (bottom) .. controls (1.5, -1) and (1.5, 1) .. (top);
					\draw[decorate, decoration={zigzag, amplitude=2pt, segment length=5pt}, thick] 
					(bottom) -- (top);
					\draw[thick] (top) -- (i0) node[midway, above right] {\Large$\mathcal{I}^+$};
					\draw[thick] (bottom) -- (i0);
					
					\draw[pattern={Lines[angle=20, distance=2pt]}]  
					(bottom) .. controls (1, -1) and (1, 1) .. (top)
					.. controls (2, 1) and (2, -1) .. (bottom);
					
					\fill[cyan!30, opacity=0.5] 
					(bottom) .. controls (1, -1) and (1, 1) .. (top)
					.. controls (2, 1) and (2, -1) .. (bottom);
					\draw[blue, thick] (bottom) .. controls (1, -1) and (1, 1) .. (top);
					\draw[blue, thick] (bottom) .. controls (2, -1) and (2, 1) .. (top);
					\node [right] at (3,0) {\Large$i^0$};
					
				\end{tikzpicture}
	}}}\\
	\vspace{1mm}
	\hspace{0.1em}
	\subfloat[]{\scalebox{0.365}{
			\begin{tikzpicture}[scale=2]
				\coordinate (Vertex) at (0,0);
				\coordinate (TopRight) at (3, 3);
				\coordinate (BotRight) at (3, -3);
				
				\fill[cyan!30, opacity=0.5] (Vertex) -- (TopRight) -- (BotRight) -- cycle;
				
				\draw[thick, black] (0,-3.5) -- (0,3.5);
				\draw[thick, black] (3,-3.5) -- (3,3.5);
				
				\draw[blue, thick] (Vertex) -- (TopRight) node[midway, above, sloped] 
				{\Large$r=0$};
				\draw[blue, thick] (Vertex) -- (BotRight) node[midway, below, sloped] 
				{\Large$r=0$};
				\draw[blue, thick] (TopRight) -- (BotRight) node[midway, above, sloped] {\Large$r=\infty$};
				
			\end{tikzpicture}
	}}
	\hspace{4.1em}
	\subfloat[]{\scalebox{0.365}{
			\begin{tikzpicture}[scale=2]
				\draw[thick] (0, -3.5) -- (0, 3.5);
				\draw[thick] (3, 3.5) -- (3, -3.5);
				
				\coordinate (BlueApex) at (1.5, 0);
				\coordinate (BlueRightTop) at (3, 1.5);
				\coordinate (BlueRightBot) at (3, -1.5); 
				
				\fill[cyan!30, opacity=0.5] (BlueApex) -- (BlueRightTop) -- (BlueRightBot) -- cycle;
				
				\draw[blue, thick] (BlueApex) -- (BlueRightTop) node[midway, above, sloped] {\Large$\rho=\kappa$};
				
				\draw[blue, thick] (BlueApex) -- (BlueRightBot) node[midway, below, sloped] {\Large$\rho=\kappa$};
				
				\draw[blue, thick] (BlueRightTop) -- (BlueRightBot) node[midway, above, sloped] {\Large$\rho=\infty$};
				
			\end{tikzpicture}
	}}
	\hspace{2.8em}
	\subfloat[]{\scalebox{0.365}{
			\begin{tikzpicture}[scale=2]
				\def\h{3.5} 
				\def\w{1.5}
				
				\fill[cyan!30, opacity=0.5] (-\w, -\h) rectangle (\w, \h);
				
				\draw[blue, thick] (-\w, \h) -- (-\w, -\h) node[midway, below, sloped] {\Large$R=-\infty$};
				\draw[blue, thick] (\w, \h) -- (\w, -\h) node[midway, above, sloped] {\Large$R=\infty$};
			\end{tikzpicture}
	}}
	\caption{Penrose diagrams showing the emergence of ${\rm AdS}_2\times {\rm S}^2$ from scalings of RN. 
		Top row: from left to right, we have extreme RN (ERN), sub-extreme RN (sub-ERN), and super-extreme RN (super-ERN). In each case, assuming (near)-extremality, the regions having an ${\rm AdS}_2$ geometry are displayed hatched.
		Bottom row: a global ${\rm AdS}_2$ is drawn and in each case the patch covered by the corresponding scaling coordinates is displayed shaded, that is from left to right, the shaded blue regions are covered by Poincare, Rindler, and global coordinates. 
	}\label{Fig:ThreeLimits}
\end{figure}

\textopenbullet\ Limit 1: ERN $\rightarrow$ Poincare ${\rm AdS}_2\times {\rm S}^2$. Depicted in panels (a) and (d) in Fig.~\ref{Fig:ThreeLimits}.

Beginning with an ERN and defining scaling coordinates as follows,
\begin{align}\label{ExtremeScalingCoords}
	Q=M \qquad \text{and} \qquad 
	r=\frac{\hat{r}-M}{\lambda M}\,,\quad t=\frac{\lambda \hat{t}}{M}
\end{align}
one finds that, in the $\lambda\to 0$ limit, ${\rm AdS}_2$ arises in Poincare coordinates:
\begin{align}\label{ExtremeRNseries}
	\frac{1}{M^2}{\rm d}s^2=-r^2 {\rm d}t^2+{{\rm d}r^2\over r^2}+{\rm d}\Omega^2+\lambda\, h_{\mu\nu}{\rm d}x^\mu {\rm d}x^\nu +\mathcal{O}(\lambda^2)
\end{align}
with the first order correction to the size of the transverse ${\rm S}^2$ measured by
\begin{align}
	h_{\theta\theta}=2r\,.
\end{align}
From the point of view of ${\rm AdS}_2\times {\rm S}^2$, within spherical symmetry, the metric component $h_{\theta\theta}$ is a gauge-invariant perturbation variable and its ${\rm SL}(2)$ orbit is given by\footnote{The remaining components of $h_{\mu\nu}$ may be found in \cite{Hadar:2020kry}, where it is also shown that \emph{any} spherically symmetric perturbation is fixed (up to gauge transformations) by the constants $(a,b,c)$ with varying $\mu$.}
\begin{align}\label{mu=0 pertn}
	h_{\theta\theta}=ar+brt+cr\left(t^2-{1\over r^2}\right)\,, \quad \text{with} \quad \mu=b^2-4ac=0\,.
\end{align}
Backreacting any perturbation characterized by \eqref{mu=0 pertn}, onto ${\rm AdS}_2\times {\rm S}^2$ produces an ERN. Notice that this backreaction modifies the ${\rm AdS}_2$ boundary conditions to produce the asymptotically flat ones of ERN. This phenomenon was dubbed \emph{anabasis} in \cite{Hadar:2020kry}.

\textopenbullet\ Limit 2: sub-ERN $\rightarrow$ Rindler ${\rm AdS}_2\times {\rm S}^2$. Depicted in panels (b) and (e) in Fig.~\ref{Fig:ThreeLimits}.

Beginning with a sub-ERN, with $Q/M<1$ parameterized by $\kappa$, writing
\begin{align}
	Q=M\sqrt{1-\kappa^2\lambda^2} \qquad \text{and} \qquad 
	\rho=\frac{\hat{r}-M}{\lambda M}\,,\quad \tau=\frac{\lambda \hat{t}}{M}
\end{align}
${\rm AdS}_2$ now arises, in the $\lambda\to 0$ limit, in Rindler coordinates:
\begin{align}
	\frac{1}{M^2}{\rm d}s^2
	&=-(\rho^2-\kappa^2) {\rm d}\tau^2+{{\rm d}\rho^2\over \rho^2-\kappa^2}+{\rm d}\Omega^2+\lambda\, h_{\mu\nu}{\rm d}x^\mu {\rm d}x^\nu +\mathcal{O}(\lambda^2) \\
	&=-r^2 {\rm d}t^2+{{\rm d}r^2\over r^2}+{\rm d}\Omega^2+\lambda\, h_{\mu\nu}{\rm d}y^\mu {\rm d}y^\nu +\mathcal{O}(\lambda^2) \nonumber
\end{align}
and the radius of the transverse ${\rm S}^2$ is corrected by
\begin{align}\label{hthth=2rho}
h_{\theta\theta}=2\rho=-2\kappa r t
\end{align}
In the above we have used the transformation between Rindler $(\tau,\rho)$ and Poincare $(t,r)$ coordinates 
\begin{align}
	\rho=-\kappa rt\,, \quad \tau=-{1\over 2\kappa}\ln\left(t^2-{1\over r^2}\right)
\end{align}
The ${\rm SL}(2)$ orbit of \eqref{hthth=2rho} is
\begin{align}
	h_{\theta\theta}=ar+brt+cr\left(t^2-{1\over r^2}\right)\,, \quad \text{with} \quad \mu=b^2-4ac=4\kappa^2>0
\end{align}
This means that any perturbation of ${\rm AdS}_2\times {\rm S}^2$ with $\mu>0$ may be thought of as an anabasis perturbation that produces, via backreaction, a sub-ERN with a near-extreme charge to mass ratio $Q/M=\sqrt{1-\mu/4}$ \cite{Hadar:2020kry}.

\textopenbullet\ Limit 3: super-ERN $\rightarrow$ Global ${\rm AdS}_2\times {\rm S}^2$. Depicted in panels (c) and (f) in Fig.~\ref{Fig:ThreeLimits}.

In this case, begin with a super-ERN,  with $Q/M>1$ parameterized by $L$, defining
\begin{align}
	Q=M\sqrt{1+L^2\lambda^2} \qquad \text{and} \qquad 
	R=\frac{\hat{r}-M}{\lambda M}\,,\quad T=\frac{\lambda \hat{t}}{M}
\end{align}
Then, in the $\lambda\to 0$ limit, we get ${\rm AdS}_2$ in global coordinates:
\begin{align}
	\frac{1}{M^2}{\rm d}s^2
	&=-(R^2+L^2) {\rm d}T^2+{{\rm d}R^2\over R^2+L^2}+{\rm d}\Omega^2+\lambda\, h_{\mu\nu}{\rm d}x^\mu {\rm d}x^\nu +\mathcal{O}(\lambda^2)\\
	&=-r^2 {\rm d}t^2+{{\rm d}r^2\over r^2}+{\rm d}\Omega^2+\lambda\, h_{\mu\nu}{\rm d}y^\mu {\rm d}y^\nu +\mathcal{O}(\lambda^2) \nonumber
\end{align}
with the first order correction characterized by
\begin{align}\label{hthth=2R}
	h_{\theta\theta}=2R=L r +Lr\left(t^2-{1\over r^2}\right)
\end{align}
where we have used the transformation between global $(T,R)$ and Poincare $(t,r)$ coordinates 
\begin{align}
	L\, T=\arctan\left(t+{1\over r}\right)+\arctan\left(t-{1\over r}\right)\,, \quad {R\over L}={1\over 2}r+{1\over 2}r\left(t^2-{1\over r^2}\right)
\end{align}
In this case, the ${\rm SL}(2)$ orbit of \eqref{hthth=2R} is
\begin{align}
	h_{\theta\theta}=ar+brt+cr\left(t^2-{1\over r^2}\right)\,, \quad \text{with} \quad \mu=b^2-4ac=-4L^2<0
\end{align}
Thus we conclude that if we fully backreact a $\mu<0$ perturbation on ${\rm AdS}_2\times {\rm S}^2$ we will end up with a slightly super-ERN of $Q/M=\sqrt{1+\mu/4}$.

To summarize, spherically symmetric perturbations of ${\rm AdS}_2\times {\rm S}^2$ in the electrovacuum theory are classified by the ${\rm SL}(2)$-invariant quantity $\mu$, read off from the correction to the extreme radius of the ${\rm S}^2$ transverse to ${\rm AdS}_2$. The invariant $\mu$ determines the  backreaction of the perturbations: for $\mu=0$ the fully backreacted solution is the ERN black hole, while for $\mu>0$ and $\mu<0$ it is the sub-ERN and super-ERN, respectively.

From now on we set $M=1$.

\subsection{Linear Aretakis from ${\rm AdS}_2$}\label{Sec:LinearAretakisAdS}

The linear Aretakis instability of a scalar perturbation on ERN may be reproduced from a purely ${\rm AdS}_2$ calculation. Below we review the relevant analysis following closely Ref.~\cite{LMRT1212}.\footnote{Another analysis based on the properties of solutions under the ${\rm AdS}_2$ dilation symmetry is found in \cite{Gralla:2017lto}.} 
First one identifies the black hole horizon with a specific ${\rm AdS}_2$ Poincare horizon and fixes the corresponding Poincare coordinates in terms of the ERN coordinates at asymptotic infinity. 
That is to say, using Eq.~\eqref{ExtremeScalingCoords} with $\lambda=1$, we fix ${\rm AdS}_2$ Poincare coordinates $(t,r)$ in terms of ERN coordinates $(\hat{t},\hat{r})$ such that $\mathcal{H}^+$ is given by $r=0\,,\, t=+\infty$ in ${\rm AdS}_2$.\footnote{Note that the Poincare ${\rm AdS}_2$ appearing in the $\lambda$ series in \eqref{ExtremeRNseries} may also be derived, for $\lambda=1$, simply by keeping leading order terms in the $r\ll 1$ approximation of the ERN metric written in $(t,r)$ coordinates.}
Then one imposes boundary conditions on the ${\rm AdS}_2$ boundary, where the near-horizon region is glued onto the far region of the black hole exterior, such that the linear Aretakis behavior for a scalar field in ERN, as reviewed in Section \ref{Sec:LinearAretakisERN}, is observed on and outside our preferred Poincare horizon in ${\rm AdS}_2$.

\begin{figure}[h]
	\centering
	\resizebox{!}{0.3\textheight}
	{
		\begin{tikzpicture}[scale=1.25]
			\coordinate (Vertex) at (0,0);
			\coordinate (TopRight) at (3, 3);
			\coordinate (BotRight) at (3, -3);
			
			\draw[thick] (0,3.5) -- (0,-3.5);
			\draw[thick] (3,3.5) -- (3,-3.5) node[midway, above, sloped] {$U=V\,,\, r=\infty$} node[pos=0.8, right] {$\mathcal{B}$};
			
			\draw[thick] (Vertex) -- (TopRight) node[midway, above, sloped] {$U=\frac{\pi}{2}\,,\,r=0$};
			\draw[thick] (BotRight) -- (Vertex) node[midway, below, sloped] {$V=-\frac{\pi}{2}\,,\, r = 0$};
			
			\draw[thick, -Stealth] (-1.5,0) -- (-1,0.5) node[pos=0.9, below right] {$V$};
			\draw[thick, -Stealth] (-1.5,0) -- (-2,0.5) node[pos=0.9, below left] {$U$};
		\end{tikzpicture}
	}
	\caption{Penrose diagram of ${\rm AdS}_2$ marked with global coordinates $(U,V)$. 
		The left boundary is $U=V+\pi$, while the right boundary $\mathcal{B}$ is $U=V$.  
		The asymptotically flat region of ERN is glued onto the Poincare patch, given by $-\pi/2<V<U<\pi/2$, at the right boundary $\mathcal{B}$.
		(See also panels (a) and (d) in Fig.~\ref{Fig:ThreeLimits})} 
	\label{Fig:AdS2}
\end{figure}
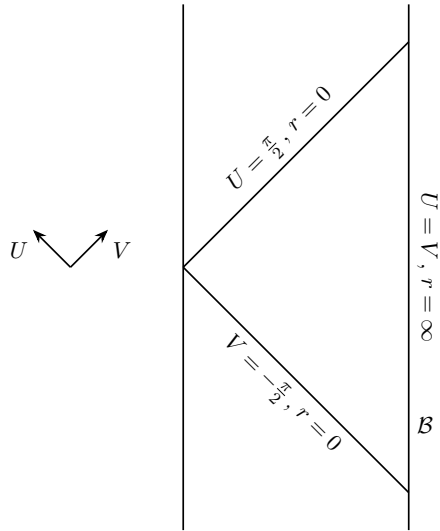

The most general solution for a neutral massless scalar $\sigma$ on ${\rm AdS}_2$,
\begin{align}\label{AdS2UV}
	{\rm d}s^2=-\frac{4 \, {\rm d}U {\rm d}V}{\sin^2(U-V)}
\end{align}
is given by 
\begin{align}\label{sigmaUV}
	\sigma=f(U)+g(V)\,,
\end{align}
for arbitrary functions $f, g$.
Here $(U,V)$ are global double null coordinates on ${\rm AdS}_2$ and they are related to Poincare coordinates with a future horizon set at $U=\pi/2$ via
\begin{align}\label{UV-to-tr}
	U=\arctan\left(t+{1\over r}\right)\,,\quad V=\arctan\left(t-{1\over r}\right)
\end{align}
In these coordinates, illustrated in Fig~\ref{Fig:AdS2}, the relevant ${\rm AdS}_2$ boundary $\mathcal{B}$, where the near-horizon region is glued onto the far region of the ERN exterior, is at $U=V$. Hence the boundary value of the scalar is
\begin{align}
	\sigma\vert_\mathcal{B}=f(U)+g(U)
\end{align}

Using regular ingoing Eddington-Finkelstein coordinates $(v,r)$, with $v=t-{1\over r}$, we see that on the future Poincare horizon at $r=0$, the scalar $\sigma$ behaves as
\begin{align}\label{sigma on H+}
	\sigma\vert_{r=0}=f(\pi/2)+g(\pi/2) -\frac{g'(\pi/2)}{v}
	+\frac{g''(\pi/2)}{2v^2} +\mathcal{O}\left(\frac{1}{v^3}\right) \,,
\end{align} 
while in the exterior, at $r=r_0>0$, it is given by
\begin{align}\label{sigma outside H+}
	\sigma\vert_{r=r_0}=\,& f(\pi/2)+g(\pi/2) -\frac{f'(\pi/2)+g'(\pi/2)}{v}
	+\frac{2f'(\pi/2)}{r_0 v^2}+\frac{f''(\pi/2)+g''(\pi/2)}{2v^2} \\ 
	&-\frac{4f'(\pi/2)}{r_0^2 v^3}-\frac{2 f''(\pi/2)}{r_0 v^3}
	+\frac{f'(\pi/2)+g'(\pi/2)}{3 v^3}
	-\frac{f'''(\pi/2)+g'''(\pi/2)}{6v^3}
	+\mathcal{O}\left(\frac{1}{v^4}\right)\,.\nonumber
\end{align}
On the other hand, using the fact that $\p_r=-\frac{\sin^2(U-V)}{2\cos^2V}\p_U$, one finds that the scalar's radial derivatives at the horizon are:
\begin{align}\label{sigma derivs on H+}
	\begin{aligned}
	\p_r \sigma\vert_{r=0}&=-\frac{1}{2}f'(\pi/2)\,,\\
	\p_r^2 \sigma\vert_{r=0}&=\frac{1}{2}f'(\pi/2)\, v+\frac{1}{4}f''(\pi/2)+\mathcal{O}\left(\frac{1}{v^4}\right)\,,\\
	\p_r^3 \sigma\vert_{r=0}&=-\frac{3}{4}f'(\pi/2)v^2-\frac{3}{4}f''(\pi/2)v+\mathcal{O}\left(1\right)
	\end{aligned}
\end{align}
Comparing Eqs.~(\ref{sigma on H+}--\ref{sigma derivs on H+}) with the linear Aretakis behavior on and outside the ERN black hole horizon $\mathcal{H}^+$, Eqs.~(\ref{linear Aretakis H},\ref{linear Aretakis no H}), we see that we get a perfect match provided we impose the conditions
\begin{align}
	\begin{aligned}
	f(\pi/2)+g(\pi/2)=0\\
	f'(\pi/2)+g'(\pi/2)=0\\
	f''(\pi/2)+g''(\pi/2)=0
	\end{aligned}
\end{align}
together with 

\textbullet\ \textit{Non-vanishing Aretakis constant}: 
$ H\equiv -\frac{1}{2}f'(\pi/2)\neq 0$, 

\textbullet\ \textit{Vanishing Aretakis constant}: 
$H\equiv -\frac{1}{2}f'(\pi/2)= 0$,  $f''(\pi/2)\neq 0$

To summarize, linear Aretakis behavior for a massless neutral scalar $\sigma$ on ${\rm AdS}_2$ is observed on a Poincare horizon ---to be identified with the future black hole horizon $\mathcal{H}^+$ in ERN--- by imposing the following boundary condition at late times on the boundary:
\begin{align}\label{linAretakisAdSbc}
	\sigma\vert_\mathcal{B},\sigma\vert_\mathcal{B}',\sigma\vert_\mathcal{B}''\to 0 \quad \text{as}\quad U\to \pi/2\,.
\end{align}
It is worth emphasizing that we are not thinking of the above as a derivation of the Aretakis instability for a scalar on ERN, but rather as a derivation of the appropriate boundary conditions to impose in a purely ${\rm AdS}_2$ analysis with respect to the preferred Poincare horizon. 
That said, we do expect it is possible to derive these boundary conditions from a full ERN analysis. Such a derivation would deal with the following two related aspects: first, the ERN ingoing Eddington-Finkelstein coordinate differs from the exact ${\rm AdS}_2$ one by a logarithmic correction that is diverging near the horizon $\hat{v}\approx v+2\ln r$; second, the relevant ${\rm AdS}_2$ for capturing near-horizon ERN physics is not the rigid one considered here but instead one that is corrected by the anabasis perturbation discussed in the previous subsection, Eq.~\eqref{mu=0 pertn}. Including the anabasis correction ensures, in the language of \cite{Hadar:2020kry}, a \emph{connected} ${\rm AdS}_2$ which maintains the necessary (minimal) information from the asymptotically flat ERN physics. Using a connected ${\rm AdS}_2$, called a \emph{near-}${\rm AdS}_2$ in JT theory, is essential for any consistent non-linear calculation which takes into account backreaction.

\section{DERN in JT}\label{Sec:DERNinJT}

We study analytically the late-time near-horizon approach to DERN by employing the two-dimensional Jackiw-Teitelboim (JT) gravity \cite{Teitelboim:1983ux,Jackiw:1984je} coupled to a massless scalar:\footnote{We are setting the two-dimensional Newton's constant to $8\pi G^{(2)}_{\rm N}=1$ and the ${\rm AdS}_2$ radius to $L=1$.}
\begin{align}\label{JTM}
	S={1\over 2}\int {\rm d}^2x \sqrt{-g} \, \Phi(R+2)-{1\over 2}\int {\rm d}^2x \sqrt{-g} \, (\nabla\sigma)^2 + {\rm bndy \,\, terms}
\end{align}
This theory describes the s-wave dynamics of the four-dimensional theory \eqref{EMS} near an ${\rm AdS}_2\times {\rm S}^2$ black hole throat close to extremality \cite{Almheiri:2014cka,Maldacena:2016upp}.
Here the matter field $\sigma$ is a direct descendant of the scalar $\phi$ in \eqref{EMS}, while the dilaton field $\Phi$ measures the variation in the area of the ${\rm S}^2$ as one climbs out of the throat. When the black hole is large, to leading order in the large area of the ${\rm S}^2$, the theory \eqref{JTM} describes accurately the non-linear dynamics near the throat.
The equations of motion are: 
\begin{gather}
	R[g] = -2\,, \qquad \Box\sigma = 0\,, \label{MatterEQ} \\
	g_{\mu\nu} \Box\Phi -\nabla_\mu\nabla_\nu\Phi -g_{\mu\nu}\Phi = \nabla_\mu\sigma\nabla_\nu\sigma-\frac{1}{2}g_{\mu\nu}\nabla_\alpha\sigma\nabla^\alpha\sigma\,. \label{PhiEQ}
\end{gather}
As a result, the metric is fixed to be always ${\rm AdS}_2$ and the matter propagates freely on it. The gravitational dynamics is therefore entirely encoded in the equation for the dilaton $\Phi$. 
Notice that in \eqref{JTM} the matter does not couple directly to the dilaton. Within spherical symmetry, this is true to leading order in the large ${\rm S}^2$ area as we expand around ${\rm AdS}_2\times {\rm S}^2$.
It means that the backreaction of the matter on the geometry is captured entirely through the matter's stress-energy source on the right hand side of the dilaton equation \eqref{PhiEQ}. This makes the theory exactly solvable.

Using the global double null coordinates $(U,V)$ from Section \ref{Sec:LinearAretakisAdS} the solution to Eqs.~\eqref{MatterEQ} is given by \eqref{AdS2UV} and \eqref{sigmaUV}, which we repeat here for convenience:
\begin{gather}
	{\rm d}s^2=-\frac{4 \, {\rm d}U {\rm d}V}{\sin^2(U-V)}\,, \qquad	\sigma=f(U)+g(V)\,.
\end{gather}
In these coordinates, Eq.~\eqref{PhiEQ} becomes:
\begin{align}\label{PhiEqs}
	\begin{aligned}
		\pdU\pdV \Phi+  \frac{2}{\sin^2\left(U-V\right)}\Phi & = 0 \,,\\
	-\frac{1}{\sin^2\left(U-V\right)}\pdU\left(\sin^2\left(U-V\right)\pdU\Phi \right) & = \mathcal{T}_{UU}\,,\\
	-\frac{1}{\sin^2\left(U-V\right)}\pdV\left(\sin^2\left(U-V\right)\pdV\Phi \right) & = \mathcal{T}_{VV}\,,
	\end{aligned}
\end{align}
where 
\begin{equation}
	\mathcal{T}_{UU} = {f'(U)}^2\,, \qquad \mathcal{T}_{VV} = {g'(V)}^2\,,
\end{equation}
are the non-vanishing components of the stress-energy tensor 
$\mathcal{T}_{\mu\nu} =\nabla_\mu\sigma\nabla_\nu\sigma-\frac{1}{2}g_{\mu\nu}\left(\nabla\sigma\right)^2$.
Using the approach of \cite{Almheiri:2014cka}, the most general solution to \eqref{PhiEqs} may be written as:
\begin{align}\label{PhiSoln1}
	\begin{aligned}
		\Phi = \Phi_{\rm vac} &-\frac{1}{\sin\left(U-V\right) } \int^{U}\dd X\sin\left(U-X \right)\sin\left(X-V \right)\mathcal{T}_{UU}(X) \\
		&+ \frac{1}{\sin\left(U-V\right) }\int^{V}\dd X\sin\left(U-X \right)\sin\left(X-V \right)\mathcal{T}_{VV}(X)\,,
	\end{aligned}
\end{align}
where
\begin{align}
	\Phi_{\rm vac} = \frac{2a\cos U\cos V + b\sin(U+V)+2c\sin U\sin V}{\sin(U-V)}\,,
\end{align}
and $a,b,c$ are integration constants. 
Here, $\Phi_{\rm vac}$ is the most general vacuum solution, and comparing notation with Section~\ref{Sec:ThreeLimits} we have 
\begin{align}
	\Phi_{\rm vac}=h_{\theta\theta}=ar+brt+cr\left(t^2-{1\over r^2}\right)\,.
\end{align}
It is instructive to rewrite the general solution in terms of the net matter flux through the ${\rm AdS}_2$ boundary $\mathcal{B}$ at $U=V$ (see Fig.~\ref{Fig:AdS2}),
\begin{align}
	\delta\mathcal{T}(U) \equiv \left(\mathcal{T}_{VV}-\mathcal{T}_{UU}\right)|_\mathcal{B} = g'(U)^2 - f'(U)^2\,,
\end{align}
with $\delta\mathcal{T}>0$ ($\delta\mathcal{T}<0$) corresponding to net flux into (out of) ${\rm AdS}_2$. In terms of $\delta\mathcal{T}$ the solution \eqref{PhiSoln1} takes the form:
\begin{align}\label{PhiSoln2}
	\Phi=\Phi_{\rm vac} + \Phi_{\rm bal}  -\frac{1}{\sin\left(U-V\right) }\int^{\pi/2}_V\dd X \sin\left(U-X \right)\sin\left(X-V \right)\delta\mathcal{T}(X)\,,
\end{align}
where 
\begin{align}
	\Phi_{\rm bal} =  -\frac{1}{\sin\left(U-V\right) }\int_V^U\dd X \sin\left(U-X \right)\sin\left(X-V \right)f'(X)^2\,.
\end{align}
Here, the `balanced' solution $\Phi_{\rm bal}$ corresponds to the case of identically vanishing net flux along the ${\rm AdS}_2$ boundary ($\delta\mathcal{T}=0$).
Notice that we have chosen the (otherwise arbitrary) upper limit of the integral appearing in \eqref{PhiSoln2} to align with the Poincare horizon at $U=\pi/2$ which we wish to identify with the future black hole horizon $\mathcal{H}^+$ of DERN. 

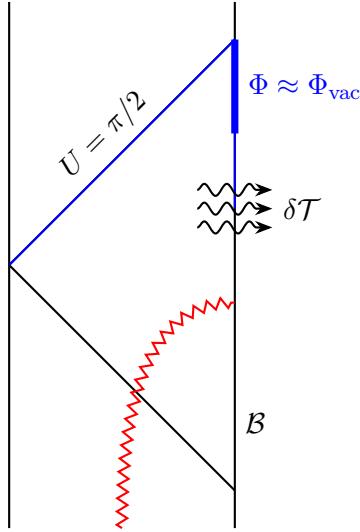
\begin{figure}[h]
	\begin{tikzpicture}[scale=1]
		\coordinate (Vertex) at (0,0);
		\coordinate (TopRight) at (3, 3);
		\coordinate (BotRight) at (3, -3);
		\draw[thick, black] (0,3.5) -- (0,-3.5) ;
		\draw[thick, black] (3,3.5) -- (3,-3.5) node[pos=0.8, right] {$\mathcal{B}$};
		\draw[thick] (Vertex) -- (TopRight) node[midway, above, sloped] {$U={\pi}/{2}$};;
		\draw[thick] (BotRight) -- (Vertex) ;
		\draw[thick, blue] (Vertex) -- (TopRight) ;
		\draw[line width=0.95mm, blue] (TopRight) -- (3,1.75) node[midway, right] {$\Phi\approx\Phi_{\rm vac}$};
		\draw[thick, blue] (TopRight) -- (3,0.75) ;
		\draw[decorate, decoration={zigzag, amplitude=2pt, segment length=5pt}, thick, red] (1.5, -3.5) to [out=90, in=180] (3, -0.5);;
		\tikzset{radiation/.style={
				decorate, 
				decoration={snake, amplitude=0.8mm, segment length=3mm, post length=1mm},
				-Stealth, 
				thick
		}}
		\draw[radiation] (2.5,0.5) -- (3.5,0.5);
		\draw[radiation] (2.5,0.75) -- (3.5,0.75)  node[right] {$\delta\mathcal{T}$};
		\draw[radiation] (2.5,1) -- (3.5,1) ;
	\end{tikzpicture}
	\caption{Penrose diagram of ${\rm AdS}_2$ with the Poincare horizon $U=\pi/2$ marked in the global coordinates $(U,V)$. There is matter leaking at early times through the right boundary $\mathcal{B}$ at $U=V$.
	At late times on $\mathcal{B}$, that is to say, along the thick blue line, the dilaton is $\Phi\approx\Phi_{\rm vac}$. 
	The asymptotically flat region of DERN is glued onto the Poincare patch at the right boundary $\mathcal{B}$, starting at late times and we can trust the gluing for as long as $\Phi\ll 1$ (\emph{i.e.} before the singularity hits $\mathcal{B}$). In particular, on the late-time thick blue portion of $\mathcal{B}$ this reduces to gluing in a static ERN exterior.
	(See also Fig.~\ref{Fig:main})} 
	\label{Fig:JTAdS2}
\end{figure}

We now need to impose the correct boundary conditions for DERN. As reviewed in Section~\ref{Sec:DERN}, DERN asymptotically has a dynamical near-horizon region which is glued onto a static ERN exterior region and the behavior of the scalar on the horizon is identical to the linear Aretakis behavior. 
This means that we are looking for non-trivial dynamical matter $\sigma$ in JT obeying the boundary conditions from Section~\ref{Sec:LinearAretakisAdS}, such that we obtain a dilaton solution $\Phi$ which reduces at late times near the ${\rm AdS}_2$ boundary $\mathcal{B}$ to an extreme vacuum one.
Therefore, first we impose on the scalar $\sigma$ the boundary conditions \eqref{linAretakisAdSbc}, which in terms of $\delta\mathcal{T}$ translate to:\footnote{Of the three conditions in \eqref{linAretakisAdSbc}, the first one does not constrain $\delta\mathcal{T}$ because it may always be achieved using the scalar's shift symmetry.}
\begin{align}\label{sigmaBC}
	\delta \mathcal{T}(\pi/2)=0\,, \qquad \delta \mathcal{T}'(\pi/2)=0.
\end{align}
Then, we note that the dilaton solution \eqref{PhiSoln2} reduces at late times on the boundary $\mathcal{B}$ to a vacuum one $\Phi\approx\Phi_{\rm vac}$ (see Fig.~\ref{Fig:JTAdS2}) and therefore from Section~\ref{Sec:ThreeLimits} we impose on the dilaton
\begin{align}\label{abcBC}
	b^2-4ac=0\,.
\end{align}
Recall that an important feature of DERN is that it lies on the threshold of black hole formation, with super-ERN and sub-ERN on either side of it. Indeed, in Section~\ref{Sec:ThreeLimits} the condition \eqref{abcBC} on $\Phi_{\rm vac}$ corresponds precisely to this threshold.

Finally, in order to be able to identify the Poincare horizon at $U=\pi/2$ with DERN's black hole horizon $\mathcal{H}^+$, we need to ensure that it is, as shown in Fig.~\ref{Fig:JTAdS2}, singularity-free. In the JT theory \eqref{JTM} the locus of the singularity is given by $1+\Phi=0$ \cite{Almheiri:2014cka}.\footnote{For the four-dimensional solution the locus $1+\Phi=0$ corresponds, in the coordinates of \eqref{RN}, to $\hat{r}=0$.} 
We find that it is not difficult to have a singularity-free horizon provided that we choose matter that is sufficiently leaky $\delta\mathcal{T}(U)<0$ at early times. In this case, the singularity does not intersect the Poincare horizon at $U=\pi/2$ and we may identify a singularity-free region in Fig.~\ref{Fig:JTAdS2} that is proximal to $U=\pi/2$ with the late-time near-horizon region proximal to $\mathcal{H}^+$ in DERN (see also the illustration in Fig.~\ref{Fig:main}). 

\begin{figure}[!h]
	\centering
	\subfloat[$H=1$\label{Fig:Heq1}]{
		\includegraphics[width=0.25\linewidth]{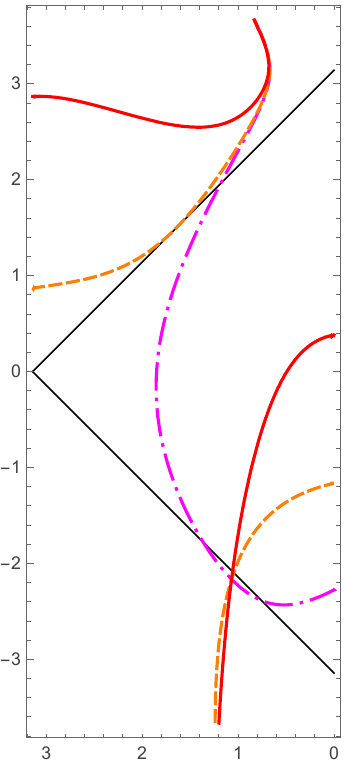}
	}
	\qquad \qquad
	\subfloat[$H=0$\label{Fig:Heq0}]{%
		\includegraphics[width=0.25\linewidth]{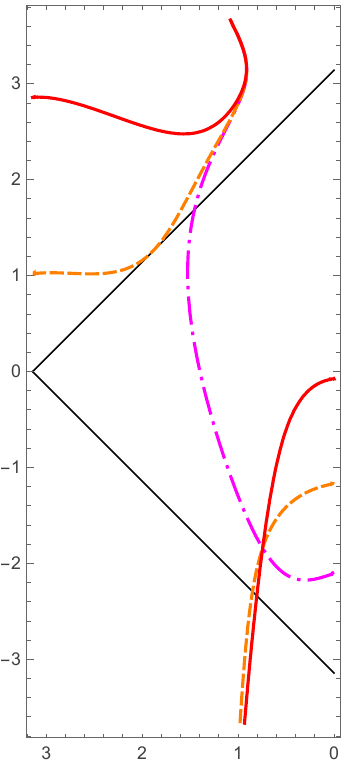}
	}
	
	\caption{Plots of the singularity locus $1+\Phi=0$ on a global ${\rm AdS}_2$ strip showing the Poincare patch $-\pi/2<V<U<\pi/2$ bounded by the solid $45^\circ$ black lines. 
	In each case, we plot the singularity curve for three different amplitudes $A$ of the corresponding net matter flux $\delta\mathcal{T}(U)<0$ leaking through the right boundary at $U=V$.  Dot-dashed magenta: $A=A_{\rm min}/10$, Dashed orange: $A=A_{\rm min}$, Solid red: $A=10A_{\rm min}$. The minimum amplitude $A_{\rm min}$ is determined in each case by the requirement that the singularity barely grazes the future horizon at $U=\pi/2$.
	For $A>A_{\rm min}$ we get a dilaton $\Phi$ that accurately describes the late-time near-horizon dynamics of a DERN.
	Both panels assume $a=2\,, b=c=0$. Left panel (a) corresponding to the dilaton \eqref{PhiSolnExampleH}: $f(U) =-2U$, $\delta\mathcal{T}(U) = -A(\pi/2-U)^2$, $A_{\rm min}=2.76$. Right panel (b) corresponding to  \eqref{PhiSolnExampleNoH}: $f(U) = (\pi/2-U)^2$, $\delta\mathcal{T}(U) = -A(\pi/2-U)^4$ \,, $A_{\rm min}=1.73$.}
	\label{Fig:PhiPlots}
\end{figure}

As a result, in the JT theory we have full analytic control of the dynamical approach to extremality: any smooth function $f(U)$ together with three constants $(a,b,c)$, subject to equation \eqref{abcBC}, and a sufficiently strong function $\delta\mathcal{T}(U)$, satisfying the conditions \eqref{sigmaBC}, give rise to a dilaton $\Phi$ in Eq.~\eqref{PhiSoln2} which describes the late-time near-horizon region of a dynamical extreme Reissner–Nordström black hole.
Many such choices lead to explicit closed form expressions for the dilaton solutions.
For example, setting $a=2\,, b=c=0$, a DERN with Aretakis constant $H=1$ may be obtained by choosing $f(U)=-2U$ and $\delta\mathcal{T}(U)=-A(\pi/2-U)^2$ with $A>A_{\rm min}\approx 2.76$, that is:
\begin{align}\label{PhiSolnExampleH}
	\begin{aligned}
		\Phi^{\rm DERN}_{H=1} = \frac{4\cos U\cos V}{\sin(U-V)} 
		+\cot (U-V) \left[2 (U-V)-\frac{A}{6} \left(\frac{\pi }{2}-V\right)^3 +\frac{A}{4}  \left(\frac{\pi}{2}-V\right)\right] \\
		-\frac{A}{4}  \frac{\sin (U) \cos (V)}{\sin(U-V)}
		+\frac{A}{4} \left(\frac{\pi }{2}-V\right)^2-2
	\end{aligned}
\end{align}
Similarly, a DERN with vanishing Aretakis constant is given by $f(U)=(\pi/2-U)^2$ and $\delta\mathcal{T}(U)=-A(\pi/2-U)^4$ with $A>A_{\rm min}\approx 1.73$, that is:
\begin{align}\label{PhiSolnExampleNoH}
	\begin{split}
		\Phi^{\rm DERN}_{H=0}&=\frac{4\cos U\cos V}{\sin(U-V)} 
		-\frac{1}{\sin (U-V)}\Biggl\{-\frac{3}{4} A  \sin(U) \cos (V) \\ 
		&\quad+\frac{A}{20} \left(\frac{\pi}{2}-V\right) 
		\Biggl[\cos (U-V) \left(2\left(\frac{\pi}{2}-V\right)^4 -10\left(\frac{\pi }{2}-V\right)^2+15\right)\\
		&\qquad\qquad-5 \sin(U-V) \left(\left(\frac{\pi}{2}-V\right)^2-3\right) \left(\frac{\pi}{2}-V\right)\Biggr]\\
		&\quad-\cos (U-V) \, (U-V)\left(\frac{2}{3} \left(U^2+UV+V^2\right)-\pi  (U+V)+\frac{\pi^2}{2}-1\right)\\
		&\quad+\sin (U-V) \left(U^2+V^2-\pi (U+V)+\frac{\pi ^2}{2}-1\right)\Biggr\}
	\end{split}
\end{align}
We note that $\delta\mathcal{T}(U)>0$, $\delta\mathcal{T}(U)=0$, or an insufficiently leaky $\delta\mathcal{T}(U)<0$ don't allow for a singularity-free horizon. In Fig.~\ref{Fig:PhiPlots} we plot the singularities associated with the two dilaton profiles \eqref{PhiSolnExampleH} and \eqref{PhiSolnExampleNoH} for varying leakage strengths $A$ and determine the quoted minimum values $A_{\rm min}$.

Using the coordinate transformation \eqref{UV-to-tr} one then obtains the near-horizon (small $r$) late-time (large $t$) solution in terms of asymptotic $(t,r)$ coordinates of our DERNs. For example, the solutions \eqref{PhiSolnExampleH} and \eqref{PhiSolnExampleNoH} imply the following late-time near-horizon approach:
	\begin{align}
		\Phi^{\rm DERN}_{H=1}\sim 2r-\frac{A r}{60t^3}\,,\qquad 
		\Phi^{\rm DERN}_{H=0}\sim 2r-\frac{Ar}{210t^5}
	\end{align}
	
It is worth emphasizing that our solutions are to be trusted only up to as early as JT is within its regime of validity in terms of describing the s-wave dynamics of the four-dimensional theory.\footnote{This means for as long as $\Phi\ll 1$ and well before approaching the singularity shown in Figs.~\ref{Fig:JTAdS2} and \ref{Fig:mainAdS}.} 
Within the regime of validity, it is interesting to ask how do our ${\rm AdS}_2$ solutions patch up to the exterior of the four-dimensional solutions. We have reviewed in Section~\ref{Sec:ThreeLimits} how the patching of the spacetime works by the time the exterior is static and used this to derive the boundary condition \eqref{abcBC}. 
At somewhat earlier times, the connection problem would involve matching the flux leaking across the ${\rm AdS}_2$ boundary and having it disperse to null infinity. This is difficult to study analytically in the full non-linear theory. 
On the other hand, if in the approach to DERN the spacetime stops evolving significantly prior to the scalar matter (as is the case in the approach to a generic sub-extreme RN \cite{Murata:2013daa}) then the problem may be studied as a linear response one. That is to say, in this case, using the results of \cite{deCesare:2024csp}, one may study analytically the spacetime perturbations around extreme RN induced by the dynamical late-time scalar and track the approach to DERN for longer than we have done here.
~\\
~\\
~\\
~

\section*{Acknowledgements}
We are grateful to A.~Castro, D.~Gajic, S.~Hadar, C.~Kehle, E.~Kiritsis, S.~Murthy, and H.~Reall for useful conversations.
This research was implemented in the framework of H.F.R.I. call “Basic
research Financing (Horizontal support of all Sciences)” under the National
Recovery and Resilience Plan ``Greece 2.0'' funded by the European Union
---NextGenerationEU (H.F.R.I. Project Number: 15384). APP is also partially
supported by UoC grant number 12030.

\bibliographystyle{utphys}
\bibliography{DERN}

\end{document}